\documentclass[draft]{agujournal2019}
\usepackage{url} 
\usepackage{lineno}
\usepackage[inline]{trackchanges} 
\usepackage{soul}

\usepackage{amsmath}
\usepackage{amssymb}
\usepackage{color}
\usepackage{tabularray}
\definecolor{MineShaft}{rgb}{0.2,0.2,0.2}

\draftfalse

\journalname{Geophysical Research Letters}

\begin{document}

%
%


\title{Data-Driven Probabilistic Air-Sea Flux Parameterization}

%
%




\authors{Jiarong Wu\affil{1}, Pavel Perezhogin\affil{1}, David John Gagne\affil{2}, Brandon G. Reichl\affil{3}, Aneesh C. Subramanian\affil{4}, Elizabeth J. Thompson\affil{5}, and Laure Zanna\affil{1}} 

\affiliation{1}{Courant Institute of Mathematical Sciences, New York University, New York, NY, USA}
\affiliation{2}{National Center for Atmospheric Research, Boulder, CO, USA}
\affiliation{3}{NOAA – Geophysical Fluids
Dynamics Laboratory, Princeton, NJ, USA}
\affiliation{4}{Department of Atmospheric and Oceanic Sciences, University of Colorado Boulder, Boulder, CO, USA}
\affiliation{5}{NOAA Physical Sciences Lab, Boulder, CO, USA}





\correspondingauthor{Jiarong Wu}{jiarong.wu@nyu.edu}



\begin{keypoints}
\item We propose a probabilistic air-sea turbulent momentum and heat flux algorithm based on neural networks trained on in-situ observations.
\item Our algorithm quantifies the uncertainty (variability) around the mean, with a mean prediction similar to existing bulk algorithms.
\item Marked seasonal sensitivities are found in deterministic and stochastic tests performed on forced single-column upper-ocean model GOTM.
\end{keypoints}


%
%

%
%


\begin{abstract}
Accurately quantifying air-sea fluxes is important for understanding air-sea interactions and improving coupled weather and climate models. This study introduces a probabilistic framework to represent the highly variable nature of air-sea fluxes, which is missing in deterministic bulk algorithms. Assuming Gaussian distributions conditioned on the input variables, we use artificial neural networks and eddy-covariance measurement data to estimate the mean and variance by minimizing negative log-likelihood loss. The trained neural networks provide alternative mean flux estimates to existing bulk algorithms, and quantify the uncertainty around the mean estimates. A stochastic parameterization of air-sea turbulent fluxes can be constructed by sampling from the predicted distributions. Tests in a single-column forced upper-ocean model suggest that changes in flux algorithms influence sea surface temperature and mixed layer depth seasonally. The ensemble spread in stochastic runs is most pronounced during spring restratification.
\end{abstract}

\section*{Plain Language Summary}
Understanding the exchange of heat and momentum between the ocean and the atmosphere is key to improving weather and climate predictions. However, quantifying these exchanges (fluxes) is difficult, and models rely on simplified equations to compute the flux values based on state variables (wind speed, air temperature, sea surface temperature, etc.). This study introduces a new statistical method based on machine learning (artificial neural networks) trained using direct field observations of fluxes to better estimate these ocean-atmosphere fluxes. The model provides not only a better estimate of the average flux, but also an understanding of its uncertainty. Preliminary tests show that this new flux estimation method has a considerable effect on the simulated state of the upper ocean, especially during certain seasonal shifts. This approach helps improve the accuracy of air-sea flux estimates and can eventually lead to better coupled weather and climate models.

%
%

\section{Introduction}
The atmosphere and ocean exchange mass, momentum, and heat across the air-sea interface. These fluxes influence ocean and atmosphere processes across a vast range of scales. However, quantifying air-sea fluxes is challenging, and significant uncertainties remain \cite{cronin_air-sea_2019}. 

In this work, we focus on momentum flux (denoted as $\tau_x$ and $\tau_y$), and on turbulent heat flux (THF), which is a non-radiative heat flux consisting of sensible heat flux ($Q_S$) due to air-sea temperature difference and latent heat flux ($Q_L$) due to air-sea humidity difference. 
Direct in-situ observations of these turbulent fluxes from atmosphere to ocean rely on measuring the covariances of turbulent fluctuations:
\begin{equation}\label{eqn:covariance}
    \tau_x = - \rho_a \overline{u'w'}, \; \tau_y = - \rho_a \overline{v'w'}, \; Q_S = -\rho_a c_p \overline{w'T'}, \; Q_L = -\rho_a L_e\overline{w'q'}. 
\end{equation}
Here $u'$, $v'$, $w'$, $T'$, and $q'$ are fluctuations of three velocity components, potential temperature, and specific humidity; $\rho_a$ is the air density; $c_p$ is the specific heat capacity at constant pressure, and $L_e$ is the latent heat of evaporation. The measurements are taken in the atmospheric surface layer, where these fluxes are assumed to be constant \cite{fairall_bulk_1996}. 

The challenges of quantifying these fluxes lie in both observational and modeling aspects. 
Due to the requirements of sophisticated instruments and careful quality control, such direct measurements are usually carried out on designated research cruises \cite{bradley_guide_2006}. Less equipped measuring platforms and remote sensing 
measure atmospheric and oceanic surface variables (wind speed, temperature, humidity, etc.) and rely on a bulk flux algorithm to compute the fluxes from the mean observed variables. The same algorithm is used in coupled weather and climate models to compute fluxes based on prognostic state variables of the oceanic and atmospheric surfaces.

The widely used flux algorithms are termed bulk algorithms because they use bulk quantities from the surface layer to model surface fluxes. The momentum and turbulent heat fluxes are formulated as proportional to the magnitude of wind speed $|U_a|$ and air-sea difference in velocity, temperature, and humidity
\begin{equation}\label{eqn:bulk}
\tau_x = \rho_a C_D S (U_a - U_o), \;   Q_S = \rho_a c_p C_H S (T_a - T_o), \; Q_L = \rho_a L_e C_E S (q_a - q_s),
\end{equation}
where $C_D$, $C_H$, and $C_E$ are the transfer coefficients for momentum, sensible, and latent heat, respectively. $S$ is the scalar wind speed relative to the ocean surface (subject to gustiness correction). Here, $\tau_x$ is aligned with the surface wind and the cross-wind component $\tau_y$ is assumed to be zero. The transfer coefficients are calculated based on the Monin-Obukhov similarity theory, where certain parameters (stability function and roughness length) are empirically determined from observations; see a complete description in, e.g., \citeA{fairall_bulk_2003} for COARE algorithm.

Currently, there are different bulk algorithms, fitted to different sets of observations \cite{brunke_which_2003, biri_airseafluxcode_2023}. Varying levels of simplifications and empirical corrections (e.g., cool-skin warm-layer or gustiness corrections) also exist. This source of uncertainty affects flux estimation in general circulation models (GCMs) and flux products. Sensitivity studies have shown that changes in bulk algorithms can considerably affect atmospheric dynamics \cite{hsu_ocean_2022, harrop_role_2018, polichtchouk_zonal-mean_2016} and oceanic state \cite{bonino_bulk_2022}. Flux products also suffer from uncertainty in bulk algorithms, in addition to uncertainty in bulk inputs \cite{yu_global_2019}.
Due to the prohibitive computational cost of high-fidelity numerical simulations, eddy-covariance (EC) measurements remain our best ``ground-truth'' for calibrating bulk algorithms, despite their sparsity and intrinsic measurement uncertainty \cite{gleckler_uncertainties_1997}. In this study, we use a quality-controlled research vessel cruise dataset provided by NOAA Physical Sciences Lab (PSL) to develop an alternative data-driven flux algorithm. 



Another potential drawback of deterministic bulk algorithms is the lack of variability. Bulk algorithms are simplified representations of the underlying dynamical processes and, at best, represent a statistically averaged value of fluxes given the observables. 
There is a significant spread of EC flux data around the prediction of bulk algorithms.
Such deviations from the mean flux values may be crucial for modeling processes on small and fast scales \cite{nuijens_air-sea_2024}; they can also have a rectifying effect for large-scale dynamics in nonlinear GCMs. More complex formulas such as sea-state-dependent parameterization \cite{edson_exchange_2013,bouin_wave-age-dependent_2024} have been proposed that incorporate additional sources of variability, but have not yet been fully validated and widely adopted.

We aim to quantify the uncertainty (variability) around deterministic flux algorithms in EC data while taking an agnostic perspective on its sources. This is achieved by applying a framework based on conditional parametric probability distributions \cite{nix_estimating_1994, barnes_adding_2021, guillaumin_stochasticdeep_2021, schreck_evidential_2024}. 
In particular, we use artificial neural networks (ANNs) to approximate the distribution parameters. Recent advances in machine learning have led to its applications in surface flux parameterization \cite{bourras_nonlinear_2007, leufen_calculating_2019, mccandless_machine_2022, cummins_reducing_2024, zhou_physical-informed_2024}, although mainly for point estimates.
In most of the paper, we use the terms ``uncertainty" and ``variability" interchangeably to refer to the spread in observed data, but its attribution will be discussed in the end. 

An estimate of uncertainty around deterministic bulk algorithms will benefit uncertainty quantification in downstream applications such as flux products and stochastic air-sea flux parameterization. Stochastic parameterizations have been shown to improve the representation of low-frequency variability and to reduce bias in the mean state \cite{williams_climatic_2012, andrejczuk_oceanic_2016, juricke_stochastic_2017, berner_stochastic_2017}. In particular, air-sea fluxes are expected to play a significant role in the upper ocean. However, the sensitivity of the upper ocean states to uncertainty in the flux algorithm is not yet well quantified. We implement our probabilistic flux algorithm in the single-column General Ocean Turbulence Model (GOTM, \citeA{umlauf_second-order_2005}) to study the effects on the upper ocean states.



\section{Data and model} \label{sec:data-model}
\subsection{Direct in-situ covariance measurements}

\begin{figure}
    \centering
    \includegraphics[width=1\linewidth]{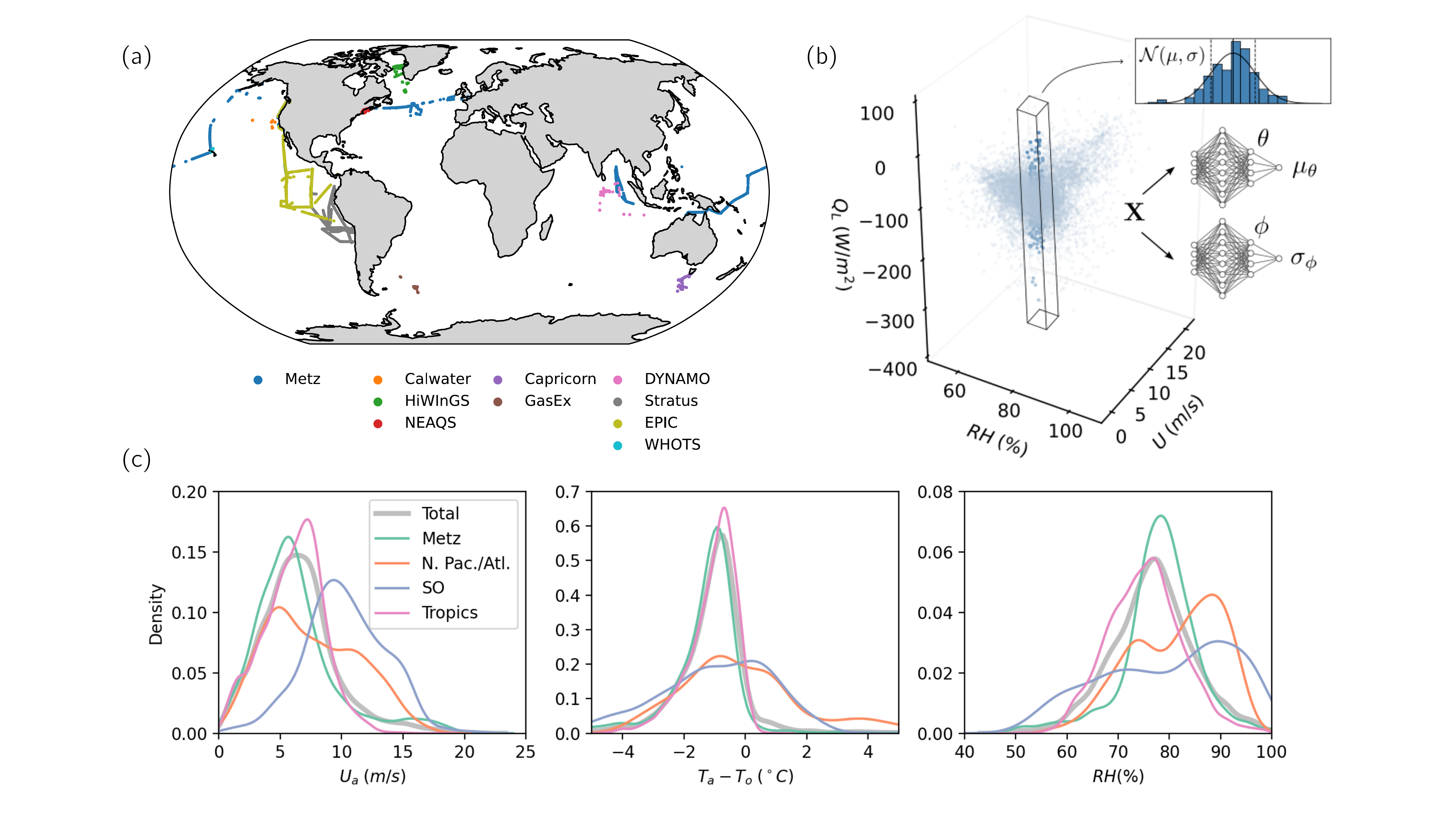}
    \caption{(a) Ship trajectories of the various cruises in the NOAA PSL dataset. (b) An illustration of the ANN-based conditional Gaussian probabilistic model. Note that we are visualizing only two of the input space dimensions, for the purpose of showing the concept of a conditional Gaussian distribution. (c) Distributions of input variables for different subsets of data. }
    \label{fig:data-model}
\end{figure}

The air-sea flux observation dataset we use is collected by decades of research cruises conducted by NOAA PSL. There are about 10,000 samples after quality control, and all are ship-borne, hourly-averaged covariance measurements according to Equation \ref{eqn:covariance}. The trajectories of the various cruises are shown in Figure \ref{fig:data-model}(a). Among the labeled cruises, Metz is a compilation of multiple earlier cruises in the 1990s; WHOTS, EPIC, DYNAMO are conducted in the tropics and Stratus in the subtropics; Calwater, NEAQS, HiWInGS in mid/high latitudes of the northern hemisphere, and Capricorn, GasEX in the southern ocean. Overall, this dataset covers various geographical locations, although a disproportional amount of data is collected in the tropics (more than 50\%). 
We note that analyzing the measurements taken along transects as time series will likely reveal additional embedded information, but in this study we treat them as independent samples of the underlying conditional distribution. 

\subsection{Probabilistic model} \label{sec:prob-model}
We consider observed bulk variables $\mathbf{X}$ (such as wind speed, humidity, etc.), unobserved variables $\mathbf{Z}$ (for example sea state, vertical wind profile, etc.), and targeted flux outputs $\mathbf{Y}=({\tau_{x}, \tau_{y}, Q_{S}, Q_{L}})$. Due to the unobserved variables $\mathbf{Z}$, the relation $\mathbf{Y} = \mathbf{F}_{model}(\mathbf{X})$ is not deterministic even when the physical laws governing the fluxes $\mathbf{Y} = \mathbf{F}_{true}(\mathbf{X},\mathbf{Z})$ are deterministic. 
Using the proposed probabilistic model, we account for all uncertainties in the data, and further attributions are discussed in Section \ref{sec:conclusion}.



We assume that each flux component $y$ of $\mathbf{Y}$ follows a uni-variate conditional Gaussian distribution with mean $\mathbf{\mu}(\mathbf{X})$ and standard deviation (std) $\mathbf{\sigma}(\mathbf{X})$:
\begin{equation}
    y \sim \mathcal{N}(\mu(\mathbf{X}),\,\sigma^{2}(\mathbf{X})).
\end{equation}
In other words, $\mu(\mathbf{X})$ is our best unbiased estimation of the flux component $y$ given $\mathbf{X}$, and the errors are distributed according to $\mathcal{N}(0,\,\sigma^{2}(\mathbf{X}))$ when conditioned on $\mathbf{X}$. This idea of conditional Gaussian distribution is demonstrated in Figure \ref{fig:data-model}(b) with the latent heat flux data. 



Having chosen the parametric distribution, we aim to learn the parameters $\mu_{\theta} (\mathbf{X})$ and $\sigma_{\phi}^2(\mathbf{X})$ with the given data. Here ${\theta}$ and ${\phi}$ denote the learnable parameters in the data-driven models for $\mu$ and $\sigma^2$.
The optimization procedure is based on computing the probability as a function of parameters (i.e., likelihood) of observing a sample value $y$ given the sample input features $\mathbf{x}$:
\begin{equation}
p (y|\mathbf{x}, \theta, \phi) = \frac{1}{\sqrt{2\pi\sigma_{\phi}^2(\mathbf{x})}}\exp{\left[-\frac{(y-\mu_{\theta}(\mathbf{x}))^2}{2\sigma_{\phi}^2(\mathbf{x})}\right]}.
\end{equation}
The optimal model parameters $\theta$ and $\phi$ minimize this negative log-likelihood loss, computed on the training dataset summed over a total of $N_\text{sample}$ \cite{nix_estimating_1994}: 
\begin{equation} \label{eqn:nllloss}
     L_\text{nll}(\theta,\phi) = \sum_{m=1}^{N_\text{sample}} \frac{1}{2} \Big [ \log({\sigma_{\phi}^2(\mathbf{x}_m)}) + {\frac{(y_{m}-\mu_{\theta}(\mathbf{x}_m))^2}{\sigma_{\phi}^2(\mathbf{x}_m)}}  \Big ] + \mathrm{const.}.
\end{equation}




\subsection{ANN architecture and training}
The mathematical framework described in Section \ref{sec:prob-model} is general and applies to any data-driven model that approximates parametric distributions.
In particular, as shown in Figure \ref{fig:data-model}(b), we use two ANNs to represent the mean $\mu_\theta$ and the variance $\sigma^2_{\phi}$ for each flux component. In this case, ${\theta}$ and ${\phi}$ are the weights and biases of the ANNs. The nonlinear activation function of the hidden layers is the Sigmoid function. The positivity of the predicted variance $\sigma^2_{\phi}$ is ensured with the exponential activation function in the last ANN layer. 

The input variables are selected based on a balance between expressibility and the risk of overfitting, as well as practical considerations for the typically available variables in GCMs. After testing different numbers of inputs and their combinations, we have chosen five inputs
\begin{equation}
    \mathbf{X}=(U_a, T_\mathrm{a}, T_\mathrm{o}, RH, p_a)
\end{equation}
which are wind speed (measured between 16 to 21 m), atmospheric surface temperature (measured between 12 to 19.5 m), sea surface temperature (measured by sea snake at 0.05 m depth), relative humidity (measured between 12 to 19.5 m), and atmospheric pressure at sea surface. Using relative humidity instead of specific humidity gives better behaviors during training, although the two can be converted given atmospheric temperature and pressure. We do not include the heights at which the meteorological variables are measured, although they are used as inputs to existing bulk algorithms. This is mainly because the ship-borne sensors are placed at a height where the vertical gradient of the atmospheric surface variables is vanishing and the height dependence cannot be explicitly learned. This is not a problem for climate-scale GCMs, which have a coarse vertical resolution, but it is a potential caveat when applied to low-lying platforms, such as buoys.

The training of ANNs is described in detail in the supplementary information. Briefly summarized here, we first train the networks on mean-square error loss and then on negative log-likelihood loss (Equation \ref{eqn:nllloss}). This two-step procedure allows the mean ANN to capture more variability in the data. The small data limit and the distribution shift shown in Figure \ref{fig:data-model}(c) are challenging for data-driven models, which we overcome through input feature selection, model design, training strategies (such as early stopping), and cross-validation. 

Having learned the parametric distribution of the fluxes for the given inputs, predictions can be made in two ways: using the mean $\mu_{\theta}$ for deterministic predictions or sampling from the distribution $\mathcal{N}(\mu_\theta, \sigma^2_{\phi})$ for stochastic predictions. In Section \ref{sec:deterministic}, we evaluate the statistical scores of the deterministic predictions. In Section \ref{sec:structure}, we analyze the predicted dependence of $\mu_{\theta}$ and $\sigma^2_{\phi}$ on inputs. In Section \ref{sec:test}, numerical experiments using both deterministic and stochastic heat fluxes are discussed. For simplicity, we denote the mean predictions as ($\mu_{\tau_x}$, $\mu_{\tau_y}$, $\mu_{Q_S}$, $\mu_{Q_L}$) and the variance predictions as ($\sigma^2_{\tau_x}$, $\sigma^2_{\tau_y}$, $\sigma^2_{Q_S}$, $\sigma^2_{Q_L}$), omitting subscript $\theta$ and $\phi$.

\section{Evaluation of the ANN-based probabilistic air-sea flux model} \label{sec:evaluation}

\subsection{Statistical scores of the deterministic predictions} \label{sec:deterministic}
We evaluate the statistical scores of the flux predictions in terms of the root-mean-square-error RMSE, the coefficient of determination $R^2$, and bias of the estimate (compared to observations):
\begin{align}
     \text{RMSE}(\hat{y},y) &= \left(\mathbb{E}[(\hat{y}-y)^2]\right)^{1/2}, \\
     R^2(\hat{y},y) &= 1-\mathbb{E}[(\hat{y}-y)^2]/\mathrm{Var}[y], \\
     \text{Bias}(\hat{y},y) &= \mathbb{E}[\hat{y}-y]
\end{align}
Here $\hat{y}$ is the algorithm prediction and $y$ is the truth, in our case the EC measurement ($\tau_{x,c}$, $\tau_{y,c}$, $Q_{S,c}$, $Q_{L,c}$). We evaluate these metrics for the ANN-based deterministic prediction ($\mu_{\tau_x}$, $\mu_{\tau_y}$, $\mu_{Q_S}$, $\mu_{Q_L}$) compared to a baseline bulk algorithm ($\tau_{x,b}$, 0, $Q_{S,b}$, $Q_{L,b}$). In particular, we choose COARE 3.6 \cite{fairall_bulk_2003}, which is the most up-to-date version of COARE. It is worth mentioning that COARE was fitted to the same shipborne EC data we used, with additional cruises and buoy measurements. 

\begin{figure}[!h]
    \centering
    \includegraphics[width=1\linewidth]{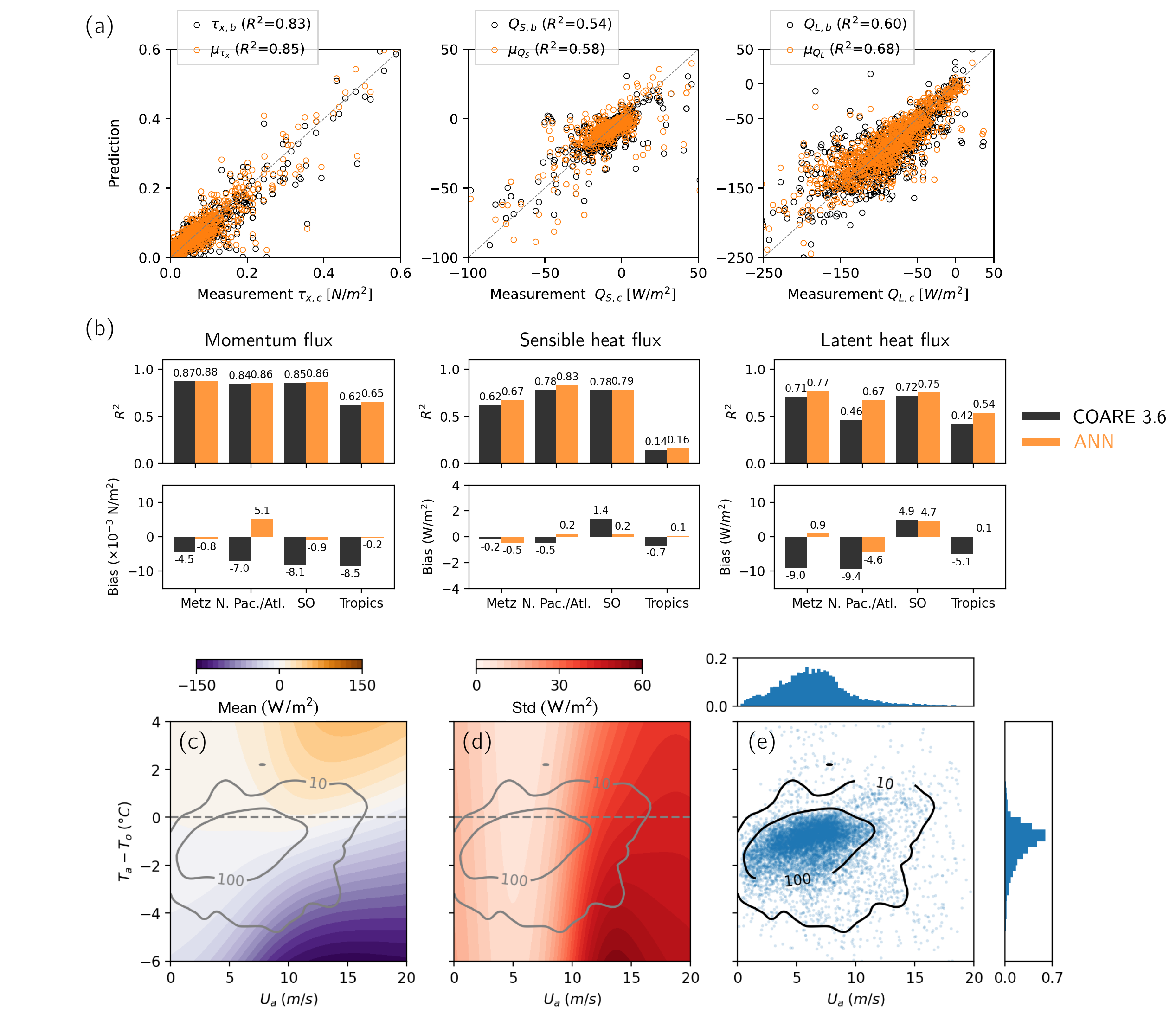}
    \caption{(a) ANN-based deterministic predictions (orange dots) and bulk algorithm predictions (black dots) plotted against measured fluxes. We are only visualizing 10\% of total samples. (b) $R^2$ and bias of ANN (orange) and bulk algorithm (black), evaluated on different geographical subsets. (c) ANN prediction of the mean of $Q_S$ on a uniform input grid. Among the five inputs, we fix sea surface temperature $T_o$ = 10 $^{\circ} C$, relative humidity $RH=80\%$, and sea level pressure $p_a$ = 1010 hPa. The dashed gray lines mark zero temperature difference $T_a$ = $T_o$. (d) ANN prediction of the std of $Q_S$ for the same grid. (e) Scatter plot of all data points and their marginal distribution. Gaussian kernel density estimation provides the gray contour lines that indicate the available samples per unit $\Delta U$ and $\Delta (T_a-T_o)$. The same contour lines are overlaid on (c,d).}
    \label{fig:pred-scatter}
\end{figure}

Figure \ref{fig:pred-scatter}(a) shows the flux predictions plotted against the EC measurements. ANN-based deterministic predictions have similar but slightly higher $R^2$ than COARE for all three fluxes, see legends in Figure \ref{fig:pred-scatter}(a), with only a subset of the input variables. For the cross-wind momentum flux $\tau_y$ (not shown here), ANN-based deterministic prediction has little predictive skills, as there is no input variable to indicate the sign of cross-wind stress. Without additional input features (such as surface wave information), it is reasonable to accept zero prediction as is in bulk algorithms. However, we can quantify the variability around the zero prediction with the current framework.

The statistical scores of both ANN and COARE vary considerably between different turbulent fluxes. The sensible heat flux $Q_S$ is the hardest to predict for both based on its lowest $R^2$ values, while the latent heat flux $Q_L$ gives the largest magnitude of error in terms of RMSE. Another observation is that scores of both ANN and COARE also vary considerably between geographical locations. Figure \ref{fig:pred-scatter}(b) shows $R^2$ and bias of three fluxes evaluated on different subsets of cruises grouped roughly by geographical locations (same as in Figure \ref{fig:data-model}). This is because each geographical subset has substantially different distributions of the input variable, as shown in Figure \ref{fig:data-model}(c). In particular, $R^2$ for sensible heat flux $Q_S$ is very low in the Tropics, only around 0.15. This is partly because the variance of the sensible heat flux is small there. Overall, the fluctuation of predictive scores over different regions is shared between ANN and COARE. 

The most salient improvement of ANN compared to baseline bulk algorithm is for latent heat flux, which is generally the least constrained of all turbulent fluxes. In certain regions, ANN sees up to around 0.2 increase in $R^2$. Overall, we notice that the estimate of latent heat given by bulk algorithm seems biased (towards predicting more evaporation from the ocean), and ANN fixes the bias to some extent, as shown in Figure \ref{fig:pred-scatter}(b). This bias correction in latent heat flux was also found in a recent paper using data-driven method \cite{zhou_physical-informed_2024} and may have important implications for global simulations. We comment that a latent heat flux model accounting for sea spray generation may offer more flexibility and help mitigate this bias \cite{andreas_improved_2015, barr_sea-state-dependent_2023}. Supplementary table S1 lists RMSE and $R^2$ evaluated on the full dataset and on different geographical subsets. The reason underlying the statistical improvement is likely that the flexible structure of ANN provides more expressibility, and therefore fits the data better where certain assumptions of traditional bulk algorithms break down, e.g., the logarithmic profile for humidity and the expression for scalar roughness lengths. This idea is further discussed in the supplementary information and ANN predictions in terms of transfer coefficients $C_D$, $C_H$, and $C_E$ are also shown (Figure S1).

\subsection{Structure of the mean and variance predicted by ANNs} \label{sec:structure}
In addition to the statistical scores, we further evaluate the ANN predictions on a reduced-dimensional uniform grid to probe and interpret the ANN flux model. This is especially important for understanding how the predicted std depends on input variables. Figure \ref{fig:pred-scatter}(c) and (d) illustrate this using sensible heat flux as an example. 

Figure \ref{fig:pred-scatter}(c) shows the prediction of $Q_S$ mean by ANN. Unlike bulk algorithms that assume $Q_{S,b} \propto S (T_a - T_o)$, the structure of ANN is more flexible, allowing non-zero flux values even when $T_a=T_o$. A similar figure for COARE is shown in supplementary Figure S2.
We did not impose any constraints on the sign of $Q_S$, contrary to \citeA{zhou_physical-informed_2024}, which used a penalty in the loss function to ensure matching signs of $T_a-T_o$ and $Q_S$. In our experience, such a penalty is not necessary for a good statistical fit. Furthermore, there is insufficient evidence for a strictly down-gradient assumption, since locally counter-gradient fluxes are possible \cite{blay-carreras_countergradient_2014, deardorff_numerical_1972} and systematic bias in measurements may exist. 

Figure \ref{fig:pred-scatter}(d) shows the prediction of std of $Q_S$ for the same grid. Uncertainty increases with wind speed, as expected, since residuals should scale with flux magnitude to some extent. A small but finite uncertainty persists at very low wind speeds, likely due to measurement challenges associated with small fluxes. For a given wind speed, the uncertainty is consistently higher for $T_a < T_o$, corresponding to unstable boundary layer conditions. As shown in Figure \ref{fig:pred-scatter}(e), data are sparse for wind speeds above 15 m/s and large temperature differences. While the ANN seems to extrapolate reasonably well, it is worth exploring methods to further constrain it where training samples are not available. We note that the same issue applies to existing bulk algorithms as well. 

\section{Idealized sensitivity tests of probabilistic air-sea flux parameterization} \label{sec:test}
To assess the impact of the proposed stochastic air–sea flux parameterization, we implement it in the single column model GOTM \cite{umlauf_second-order_2005}, which provides a controlled environment without the complex interactions with other nonlinear processes present in GCMs and significantly reduces the computational cost of running ensembles.
The governing equations are a set of 1D diffusion-type equations 
\begin{equation} \label{eqn:1D}
\partial_t u = - \partial_z \overline{w'u'} + fv, \;
\partial_t v = - \partial_z \overline{w'v'} - fu, \;
\partial_t T = - \partial_z \overline{w'T'}, \;
\partial_t S = - \partial_z \overline{w'S'}, \;
\end{equation}
where the prognostic variables are the horizontal velocity components $u$ and $v$ (subject to Coriolis force), temperature $T$, and salinity $S$. The turbulent fluxes $\overline{w'u'}$, $\overline{w'v'}$, $\overline{w'T'}$, $\overline{w'S'}$ are not resolved but are instead parameterized by vertical mixing schemes. For robustness, we test our air-sea flux algorithms in combination with two ocean surface boundary layer (OSBL) mixing parameterizations: the K-profile-parameterization (KPP) scheme \cite{large_oceanic_1994} and a more sophisticated second-order closure $k-\epsilon$ scheme \cite{umlauf_second-order_2005}, both used in GCMs but typically under different resolutions. Air-sea momentum and heat fluxes serve as boundary conditions to Equation \ref{eqn:1D}. Additionally, surface fluxes can impact the mixing parameterization. 


\subsection{THF time series at OWS Papa}
We use Ocean Weather Station (OWS) Papa (50°N, 145°W) as our test site. This long-term North Pacific station provides hourly meteorological and oceanic measurements, including temperature and salinity profiles. Its weak horizontal advection makes it a common benchmark for testing vertical mixing parameterization in single-column models \cite{umlauf_second-order_2005, li_integrating_2021}. Here, our goal is simply to use this location to demonstrate the use of stochastic flux parameterization and to assess the potential impact on upper ocean state, with given vertical mixing parameterizations. We discuss the flux predictions given the observed input variables at OWS Papa before showing results from the forced experiments. 

\begin{figure} [!h]
    \centering
    \includegraphics[width=1\linewidth]{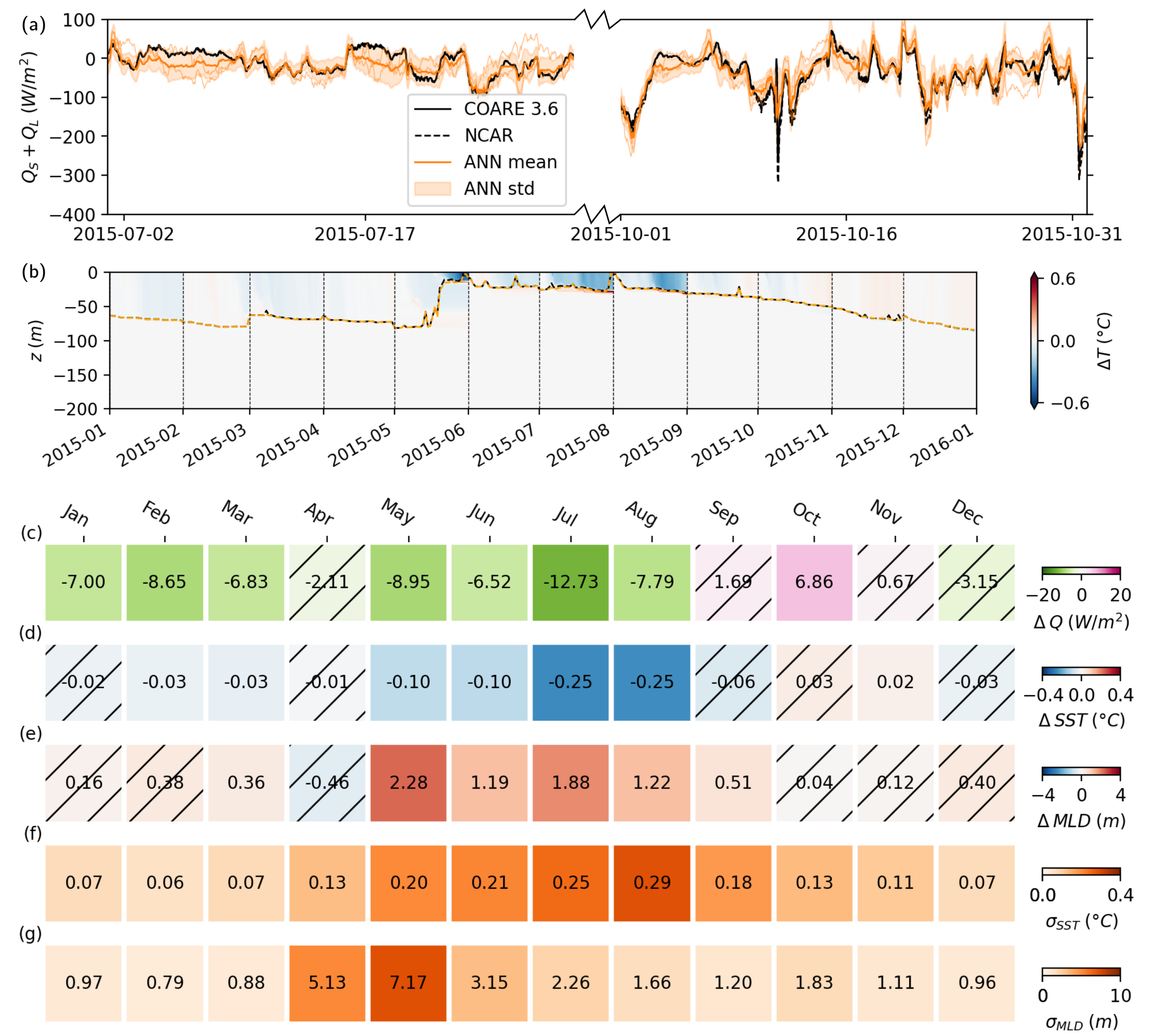}
    \caption{(a) Example THF time series for July and October 2015, computed using observed input variables at OWS Papa. Black solid line: COARE 3.6; black dashed: NCAR; orange line: ANN. Orange shade shows one standard deviation for THF predicted by ANN, which is used to generate noise-perturbed heat fluxes (examples shown by thin orange lines). (b) Difference of temperature profiles in the deterministic single-column run $\Delta T (z,t) = T_\text{ANN} (z,t) - T_\text{bulk}(z,t)$. Black and orange dashed lines show the diagnosed MLD for bulk and ANN, respectively. (c) Monthly-averaged THF discrepancies between ANN and COARE. (d) SST and (e) MLD responses in the deterministic runs between (ANN minus COARE). Trends with inconsistent signs over the years are masked with hatching. (f) Monthly-averaged SST and (g) monthly-averaged MLD spread in 20 ensemble runs. The spread is monthly averaged while the instantaneous value can be larger (see Figure S5).}
    \label{fig:ows_papa}
\end{figure}

Figure \ref{fig:ows_papa}(a) shows an example of ANN-predicted (deterministic) THF time series  $Q = \mu_{Q_S} + \mu_{Q_L}$, compared to bulk algorithms, zoomed in to July and October 2015. We plot two bulk algorithms - COARE 3.6 and NCAR \cite{large_global_2009} - to show that the ANN deviations exceed those between different bulk algorithms. As with the THF itself, the difference between the ANN and bulk predictions is seasonal, since the input distributions vary throughout the year. Figure \ref{fig:ows_papa}(c) shows the monthly averaged flux discrepancy between ANN and COARE 3.6. The months of July and October see the most negative (more heat flux out of the ocean) and positive differences. 
Notice that within a period of high THF around Oct 10, ANN predicts the least extreme negative values while NCAR predicts the most extreme negative value, reflecting all three algorithms' divergence at high wind speed, where few measurements exist to constrain them. 

Figure \ref{fig:ows_papa}(a) also shows the ANN-predicted $\pm 1\:\sigma_{Q}$ for THF. We assume no covariance between the sensible and latent heat flux residuals, so the uncertainty of their sum simply follows $\sigma_{Q}^2 = \sigma_{Q_S}^2 + \sigma_{Q_L}^2$. Importantly, the predicted $\sigma_{Q}$ is state dependent, so the uncertainty envelope changes over time. Given the ANN-predicted mean and std, we can create an ensemble of stochastically perturbed fluxes, and two examples are shown in Figure 3(a). The stochastic perturbation is expected to stem from certain unobserved processes with characteristic correlation time scales. Therefore, instead of white noise, we use temporally correlated noise $\epsilon$ generated through a discrete autoregressive model of order one - AR(1) \cite{giglio_role_2017}. 
The correlation time of 60 hours is estimated from the OWS Papa data. Note that the bulk algorithm predicted THF is generally within $\pm\sigma_Q$, and thus can be interpreted as one possible ensemble member. 

Despite relatively weak horizontal advection, there is still a net heat flux imbalance at OWS Papa \cite{large_oceanic_1994} if horizontal advection is ignored. We mitigate this by restarting the simulation with observed temperature and salinity profiles every month and by emphasizing inter-comparison between simulations rather than direct validation against observations. Details of the numerical experiments are provided in the supplementary information. 

\subsection{Sensitivity of the upper ocean state to THF uncertainty}
With the monthly restart and stochastically perturbed THF ensemble, our tests can be understood as experiments that quantify the short-term uncertainty in the upper ocean state due to uncertainty in the flux algorithm. In the following discussion, we focus on two characteristic quantities: sea surface temperature (SST) and mixed layer depth (MLD). 

At OWS Papa, the mixed layer is deep in the winter because strong winds and surface cooling induce convective mixing, and shallow in the summer because weaker winds and more radiative heating lead to stable stratification. In spring, as the radiative heating gains and the THF decreases in magnitude, there is a fast restratification process, whereas in fall, the erosion of the stratification is more gradual. This annual cycle is relatively well captured given the monthly restart, as shown in Figure S3.

Related to this annual cycle, our main finding is a significant seasonality in the sensitivity to surface heat flux. Figure \ref{fig:ows_papa}(b) shows the difference in temperature profiles between ANN and COARE flux forced runs, as well as the MLD, for 2015. There is a perceivable difference in temperature (fairly uniform across the mixed layer) as a result of a change in surface heat flux. ANN generally predicts more surface heat loss, which deepens the mixed layer and cools the sea surface. When the mixed layer is the shallowest in the summer months, there is the largest change in SST, consistent with a heat budget idea. The seasonality is confirmed by the composite statistics for four years of monthly-averaged change in SST and MLD are summarized in Figure \ref{fig:ows_papa}(d) and (e). We note that these effects are consistent across $k-\epsilon$ and KPP (see Figure S6), suggesting that the effects of air-sea fluxes is somewhat insensitive to the OSBL parameterization used. 

Perhaps more important than comparison between any two specific deterministic runs, the ensemble of stochastically perturbed simulations provides an estimate of the variability in SST and MLD induced by the variability of THF. From the 20 ensemble member runs, we do not observe systematic drift induced by the stochastic forcing; the ensemble mean of the stochastic runs closely tracks the deterministic run (see Figure S5). This is expected since we are performing forced experiments and there is no feedback of state variables (such as SST) to surface forcing. However, we observe considerable spread between ensemble members, which is summarized in Figure \ref{fig:ows_papa}(f) and (g), again with clear seasonal patterns. 
The largest spread in MLD occurs during the months of April and May, since stochasticity can significantly affect the onset of restratification. This may have important implications for biogeochemical processes, suggesting that comparisons between modeled and observed upper ocean profiles, especially during restratification, may be sensitive to surface THF uncertainty. The largest spread in SST is seen around August, again because the mixed layer is shallow and SST is sensitive to surface heat fluxes. As in deterministic runs, the effects seen in stochastic runs are not particularly sensitive to different OSBL parameterizations. 


\section{Conclusion and discussion} \label{sec:conclusion}
We propose an ANN-based probabilistic model for turbulent air-sea fluxes. State-of-the-art bulk algorithms represent the statistical mean of eddy-covariance flux observations for a given set of state variables. Our work reinforces this idea by providing an alternative mean estimate using a purely data-driven approach. The ANN-predicted mean is similar to existing bulk algorithms, with marginally higher statistical correlation to data and reduced bias for latent heat flux. Additionally, the ANN-predicted standard deviation around the mean offers a measure of uncertainty (variability). Tests using OWS Papa data and a single-column model show that the difference in THF between the different algorithms' estimates exhibits signs of seasonality, as does the sensitivity of upper ocean states, despite relatively small magnitudes. Stochastic ensemble runs indicate that the spread of MLD is largest during spring restratification, while the spread of SST is largest during summer when the mixed layer is shallow.

We comment that the current framework can also be used to estimate the standard deviation around a given mean estimation, e.g., any existing bulk algorithm, by considering the residual between the observation and the mean estimation. It is important to note that the uncertainty we estimate does not distinguish between measurement uncertainty \cite<both instrument and sampling, with sampling uncertainty being the dominant factor, see for example>{rannik_random_2016} and missing physics (the inability to account for all predictive input variables). Such a partition is difficult and likely depends on the averaging window of the EC flux, which will be examined in future studies.
Nonetheless, the ANN-estimated uncertainty highlights regions in the input space where the data is more ``scattered''. This motivates field campaigns for better sampling of these regimes and examination of measurement uncertainty, as well as the use of high-fidelity numerical simulation to constrain the flux estimate \cite{clayson_new_2023}.  

There are several possible extensions of this work in the future. The ANN model uses a limited set of input variables, which is why the improvement over the existing algorithm may be considered marginal. 
Further improvement is possible, as several additional variables are measured and available in the dataset (e.g., turbulent kinetic energy), but many of them are not yet prognostic variables in GCMs. The height dependence is difficult to learn from the current data set alone, but incorporating data from other measurement platforms (e.g., buoy) might help. We also need more high-quality vertical profile measurements of surface wind, temperature, and humidity. The probabilistic model presented assumes a univariate conditional Gaussian distribution for each flux component, which may not fully capture the underlying complexity of unobserved physical processes. The model can be extended to relax the Gaussian assumption and include the covariance between flux components. It would also be valuable to explore potential connections between the data-driven formulation and ad hoc modifications commonly used in flux algorithms, such as the gustiness factor.
Finally, the analysis at OWS Papa provides an initial estimate of the sensitivity magnitude but is unlikely to be representative of the global ocean with varying meteo-ocean conditions and coupling dynamics across regions. We plan to extend our analysis to global simulations to assess the broader impact of the proposed air-sea flux algorithm.
 

In summary, the probabilistic framework proposed in this paper provides a formal way to represent uncertainty in air–sea flux estimates, which is particularly important given the extreme scarcity of direct observations, despite their continued role as the primary means for constraining air–sea flux parameterizations.
While existing bulk algorithms represent the mean values of turbulent fluxes given limited input variables, our approach takes a step towards examining the variability around the mean values. Stochastic air-sea flux parameterization offers a promising alternative to the deterministic approach. It is crucial to prescribe the appropriate magnitude and correlation scale of noise with varying model resolutions, which needs to be better understood in future studies.  
In the single-column test, the spread induced by stochastic residuals exceeds that of different deterministic flux algorithms. Therefore, extending the testing to global GCMs could potentially affect the variability and address long-standing biases through interaction with horizontal transport and other nonlinear dynamics.

\section*{Conflict of Interest}
The authors declare no conflicts of interest relevant to this study.

\section*{Open Research Section}
The NOAA PSL data are documented and available here \url{https://downloads.psl.noaa.gov/psd3/cruises/}. The code for ANN training and evaluation, as well as the model weights are openly available \cite{jiarong_wu_2025_17554656}. The GOTM code is available here \url{https://gotm.net/portfolio/}. In particular, this work used an implementation of GOTM in the PDE solver Basilsik \cite{popinet_vertically-lagrangian_2020} \url{http://basilisk.fr/src/test/ows_papa.c}. For comparison to bulk algorithms, we used the aerobulk-python package \url{https://github.com/xgcm/aerobulk-python}.

\acknowledgments
This research received support through the NSF STC, Learning the Earth with Artificial Intelligence and Physics, LEAP (Grant number 2019625). We thank Julia Simpson, Pierre Gentine, and Bia Villas Bôas for helpful discussions and the anonymous reviewers for their useful feedback. Aakash Sane, Qing Li, and Stéphane Popinet provided valuable advice regarding the GOTM implementation. We acknowledge the continued efforts from NOAA PSL lab (including Ludovic Bariteau, Chris Fairall, and many others) in obtaining and publishing the EC datasets. The statements, findings, conclusions, and recommendations are those of the author(s) and do not necessarily reflect the views of the National Oceanic and Atmospheric Administration, or the U.S. Department of Commerce. This research was also supported in part through the NYU IT High Performance Computing resources, services, and staff expertise.


%
%

\bibliography{reference}

\end{document}